\documentclass[onecolumn,fleqn,usenatbib]{mnras}
\usepackage{booktabs}
\usepackage{graphicx}	
\newcommand{\flux}{$\rm {ergs\ s^{-1}cm^{-2}}$}
\newcommand {\kms}{${\rm km\,s}^{-1}$}

\newcommand {\HST}{{\it HST}}
\newcommand {\lya}{Ly$\alpha$}
\newcommand {\lyb}{Ly$\beta$}
\newcommand {\hb}{H$\beta$}
\newcommand {\hg}{H$\gamma$}
\def\arcsecpoint{\ifmmode ''\!. \else $''\!.$\fi}
\newcommand {\mgii}{\ion{Mg}{2} $\lambda 2798$}           
\newcommand {\civ}{\ion{C}{4} $\lambda 1549$}           
\newcommand {\cii}{\ion{C}{2} $\lambda 1335$}           
\newcommand {\ovi}{\ion{O}{6} $\lambda 1034$}           
\newcommand {\oiii}{[\ion{O}{3}] $\lambda 5007$}
\newcommand {\pair}{[\ion{O}{3}] $\lambda\lambda 4959/5007$}                    
\newcommand {\fe}{\ion{Fe}{2}}                   
\newcommand {\fee}{\ion{Fe}{3}}            
\newcommand {\uva}{\ion{Fe}{2} $\lambda 1071$}            
\newcommand {\feop}{\ion{Fe}{2}$_{\rm opt}$}                  
\newcommand {\feuv}{Fe$_{\rm UV}$}
\newcommand {\uvb}{\ion{Fe}{2}+\ion{Fe}{3} $\lambda 1123$}            
\newcommand {\he}{\ion{He}{2} $\lambda 4686$}
\newcommand {\nv}{\ion{N}{5} $\lambda 1240$}
\newcommand {\ratio}{\ion{Fe}{2}$_{\rm opt}$/H$\beta$}
\newcommand {\lm}{$\lambda$}

\usepackage{lscape}
\usepackage{longtable}

\newcounter{species} 

\def\ion#1#2{\setcounter{species}{#2}#1$\;${\scriptsize\Roman{species}}\relax}
\title[UV Fe Emission]{Far-UV Fe Emission as Proxy of Eddington Ratios}

\author[W. Zheng]{
Wei Zheng$^{1}$\thanks{E-mail: wzheng@jhu.edu}
\\
$^{1}$Department of Physics and Astronomy, Johns Hopkins University, Baltimore, MD 21218, USA  
}

\pubyear{2021}
\begin{document}
\label{firstpage}
\pagerange{\pageref{firstpage}--\pageref{lastpage}}
\maketitle

\begin{abstract}
The Eddington ratio is a key parameter that governs the diversity of 
quasar properties. It can be scaled with a strong anti-correlation 
between optical \feop\ and [\ion{O}{3}] emission.
In search of such indicators in the far-UV band, the \HST\ far-UV spectra of 150 
low-redshift quasars are analyzed in combination with their
optical SDSS counterparts. The strength of \uvb\ emission is significantly correlated with
that of \feop. A moderate correlation may also exist between \uva\ and \feop.
The finding opens the possibility
that far-UV \fe\ emission may serve as a new gauge of the Eddington ratios.
The high- and low-ionization lines in the far-UV band display different patterns: 
for the quasars with higher Eddington ratios, the low-ionization UV lines are stronger, and the 
high-ionization lines are broader and weaker. 
\end{abstract}


\section{Introduction}\label{sec_intr}

The presence of emission lines is characteristic in the 
broadband spectra of active galactic nuclei (AGN). 
The line profiles and strengths serve as essential diagnostics of the physical 
conditions in the vicinity of central black holes. 
It is generally believed that the large dispersion of line widths is mainly 
attributed to an orientation effect \citep{jarvis,shen14}, and line widths are often used 
to classify AGNs \citep{urry}.
One of the most significant correlations between line strengths is with the optical \fe\ emission:
\citet{boroson92} applied the Principal Component Analysis (PCA) to the optical spectra of 87 
low-redshift quasars ($z \la 0.5 $) to find a 
strong anti-correlation between the intensities of \fe\ \lm 4565
[Opt 37+38 multiplets \citep{vb}; \feop\ hereafter] and \oiii\ emission.
\citet{boroson} suggested that the underlying reason for this ``Eigenvalue 1''
is the Eddington ratios for these quasars: the ones with
higher Eddington ratios display stronger \feop\ and 
narrower \hb\ profiles \citep{sulentic02,marziani14}. These 
correlations between \feop/\hb\ ratios, \hb\ line widths
and Eddington ratios are supported by the work of 
\citet{shen14}, which is based on a considerably larger database of approximately 20,000 
quasars in the Sloan Digital Sky Survey \citep[SDSS,][]{york,shen11}. 

The majority of quasars are between redshifts $2-3$, and their \feop\ emission is shifted into the 
near-infrared band and beyond the SDSS scope. For these high-redshift quasars, however,
their \lya\ emission is shifted into the optical band and becomes readily observable. It 
is, therefore, interesting to search for far-UV (FUV) emission that is correlated to Eigenvalue 1.  
Several studies have characterized the UV emission lines,
exploring the trend of their properties from the optical to UV and the 
possibility of estimating the masses of central black holes by UV emission 
lines \citep{mclure,kgc,wang}. 

The UV spectra of quasars are rich in \fe\ and \fee\ emission \citep{wills,laor,vw,vb}. 
While a correlation between \fe\ features in the near-UV and optical bands is known 
\citep{sameshima, kdp, le}, no such a correlation, to our knowledge, has been established in the FUV band.
An interesting and potentially important FUV feature is the pair of \uva\ and \uvb\ emission: 
a PCA of \citet[][S06 hereafter]{suzuki} grouped 50 low-redshift quasars ($z \la 1$) into four classes 
with significantly different \lya\ emission. Note the difference in this pair of weak \fe/\fee\ 
emission: in Table 2 of S6, the average equivalent width (EW) of \uva\ is $0.9 \pm 0.9$ \AA\ and that 
for \uvb\ is $0.1 \pm 0.1$ \AA\ for Class-I quasars. As a comparison,
these values are $2.7 \pm 1.5$ and $2.2 \pm 1.1$ \AA, respectively, for Class-III quasars.
However, this work did not include optical spectra. Other studies with
measurements in both far-UV (FUV) and optical bands (in the restframe, and hereafter) 
were based on small sample sizes of $\sim 20$ quasars and with no measurements 
below the \lya\ wavelengths \citep{shang07,vmk}.

A decade after the deployment of the Cosmic Origin Spectrograph \citep[COS,][]{green},
the UV database of quasars has increased significantly. This study utilizes the 
archival spectra of 150 low-redshift SDSS quasars and compares the properties 
of FUV and optical emission lines. 
With a larger and better database, it uses optical \fe\ emission to scale its 
UV counterparts, consolidates the preliminary S06 results on \uva\ and \uvb,
and explores their potential to serve as a proxy of Eddington ratios.

\section{Data Analyses}\label{sec_data}

A sample selection of UV and optical spectra was carried out by cross-checking the SDSS quasar catalogs 
\citep{q7,q10,q12,q14} with the Hubble Space Telescope (\HST) server at the Space Telescope 
Science Institute (STScI). The search within 3\arcsec\ between the optical and UV coordinates yielded 418 
SDSS quasars of $0.05 < z < 0.8$. The selected redshift range assures a sufficient coverage of the 
redshifted  \feop\ and \oiii\ lines as well as a potential coverage of \ovi, \uva\ and 
\uvb. For simplicity, the \uvb\ emission is referred to as \feuv\ hereafter. 
The archival \HST/UV spectra are not as homogeneous as the SDSS, as they were taken with 
various spectral resolutions and wavelength coverage. 
The majority of UV spectra (318)  were obtained with the COS instrument, and 
the rest were with three other \HST\ spectrographs (FOS, HRS, and STIS). 
If a quasar has been observed with COS and other UV spectrographs
in a similar wavelength range, its COS data were used. 
If multiple observations with the same spectrograph were made, the spectra were combined 
with the weights of exposure times. 
The data were then rebinned with a pixel size of 0.5 \AA\  in the observer's frame. 
Some data were rejected because of their low signal-to-noise ratios 
(average $S/N<5$ per pixel, as calculated from the rebinned data arrays of fluxes and
errors in the fitting windows between 995 and 1170 \AA),
significantly incomplete  ({\it i.e.}, missing more than half of the emission profile) 
\ovi\ or \feuv\ profiles, or seriously contaminating broad absorption lines. 
The final sample consists of 150 quasars between $0.1 < z < 0.7$, of which 132 are with COS 
spectra. As a comparison, the FOS sample of \citet{kpd} consists of 61 quasars in the same 
redshift range, and that of S06 is 31.
The optical spectra were retrieved from the SDSS DR14 server. All data consist of 
a pair of flux and error arrays. The average redshift is 0.321, and the mean redshift 0.270.

Spectral analyses of continua and emission lines were carried out using the task 
{\tt Specfit} \citep{specfit}. 
Absorption features of ${\rm EW} > 0.1$ \AA\ were marked and excluded in the fitting wavelength windows.
For each optical spectrum, the following components were first applied: 
a power-law continuum, dual broad Gaussian components for \hb, a Gaussian component for \he.
A set of \feop\ templates of \citet{veron}, convolved with Lorentzian profiles
between 1000 and 8000 \kms, were generated and then divided into two sets by wavelengths:  one
to fit \feop\ at wavelengths $< 4700$ \AA, and another companion set of 
\fe\ templates at wavelengths $> 4800$ \AA\ to fit the \fe\ multiplets at $\sim 4900 - 5050$ \AA\ 
\citep[Opt 42;][]{vb}. {\tt Specfit} took these two sets of user-supplied components of 
velocity-broadened templates along with two free parameters: the scaling factor in 
fluxes and redshift for wavelengths. 
The scaling factors were allowed to vary independently, and 
the redshifts for these \fe\ templates were set as the systemic redshifts and  
allowed to vary slightly within the range of $|dz| < 0.005 (1+z)$. 
The best fitting results included the templates' broadening velocities, scaling factors and redshifts.

For Gaussian components, their centroid wavelengths, fluxes, and widths 
were set free, except for the \pair\ doublets: their ratios of fluxes and centroid wavelengths were tied to the 
intrinsic atomic data and their line widths were identical. 
Note that all the centroid wavelengths in {\tt Specfit} fitting were 
redshifted, but their values shown in this paper are in the restframe. 
The broad \hb\ components (also for other pairs of dual components)
were of Full Width at Half Maximum (FWHM, or ``width'')
$\ga 3000$ \kms, and $1500 - 3000$ \kms. If a spectrum  displays a discrete 
narrow \hb\ feature and the fit narrow \hb\ component showed an
${\rm FWHM} \la 1500$ \kms, an additional narrow \hb\ component of 
${\rm FWHM} < 1500$ \kms\ was added. 
Narrow components of FWHM $< 1500$ \kms\ \citep{zakamska} were excluded in 
calculating broad \hb\ fluxes as they are unlikely formed in broad-line regions.

\begin{figure}
        \includegraphics[width=\columnwidth]{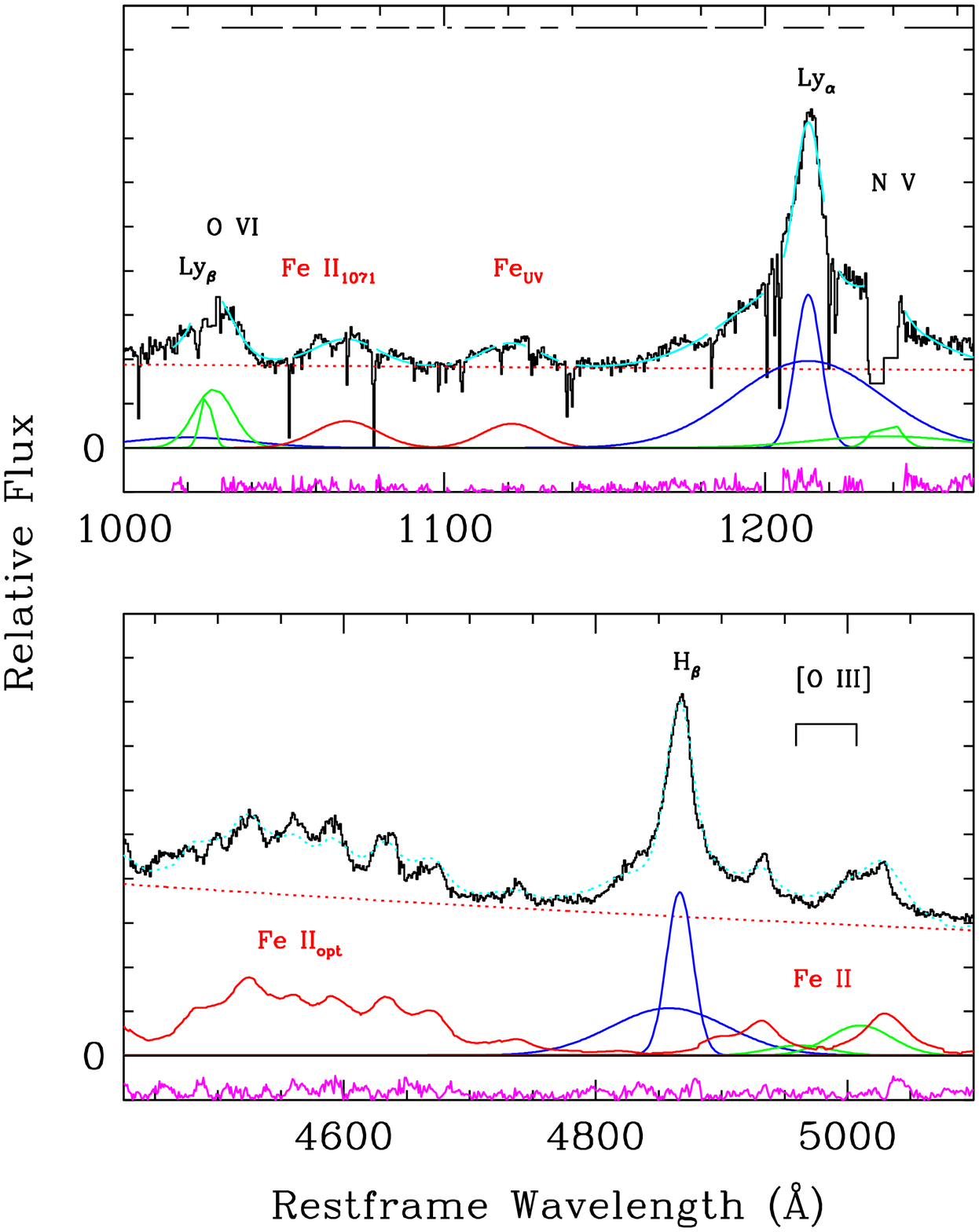}
        \caption{Spectral fit to UV and optical \fe\ features in quasar SDSS1619+2543
          ($z= 0.2685$). The UV fitting windows are marked by horizontal bars in the upper panel.
          The {\tt Specfit} fitting results are plotted in cyan, the power-law continua
          in red, and the absolute values of residues in the fitting windows
          in magenta (with a down shift to avoid overlapping).
          The \fe\ components are plotted in red, hydrogen lines in blue,
          and other metal (oxygen and nitrogen) lines in green.
          The optical \fe\ features are fit with velocity-convolved templates of
          \citet{veron}, with a
          normalization break at $\sim 4700 - 4800$ \AA,
          and the others with Gaussians. The data discontinuity near 1025 \AA\ is the result of
          geocoronal contamination, and that near 1237 \AA\ an instrumental gap.
          Note that the narrower components of \ion{O}{6} and \ion{N}{5}
          are largely outside the fitting windows; therefore their fitting results are uncertain.
        }
        \label{exam}
\end{figure}

\begin{figure}
        \includegraphics[width=\columnwidth]{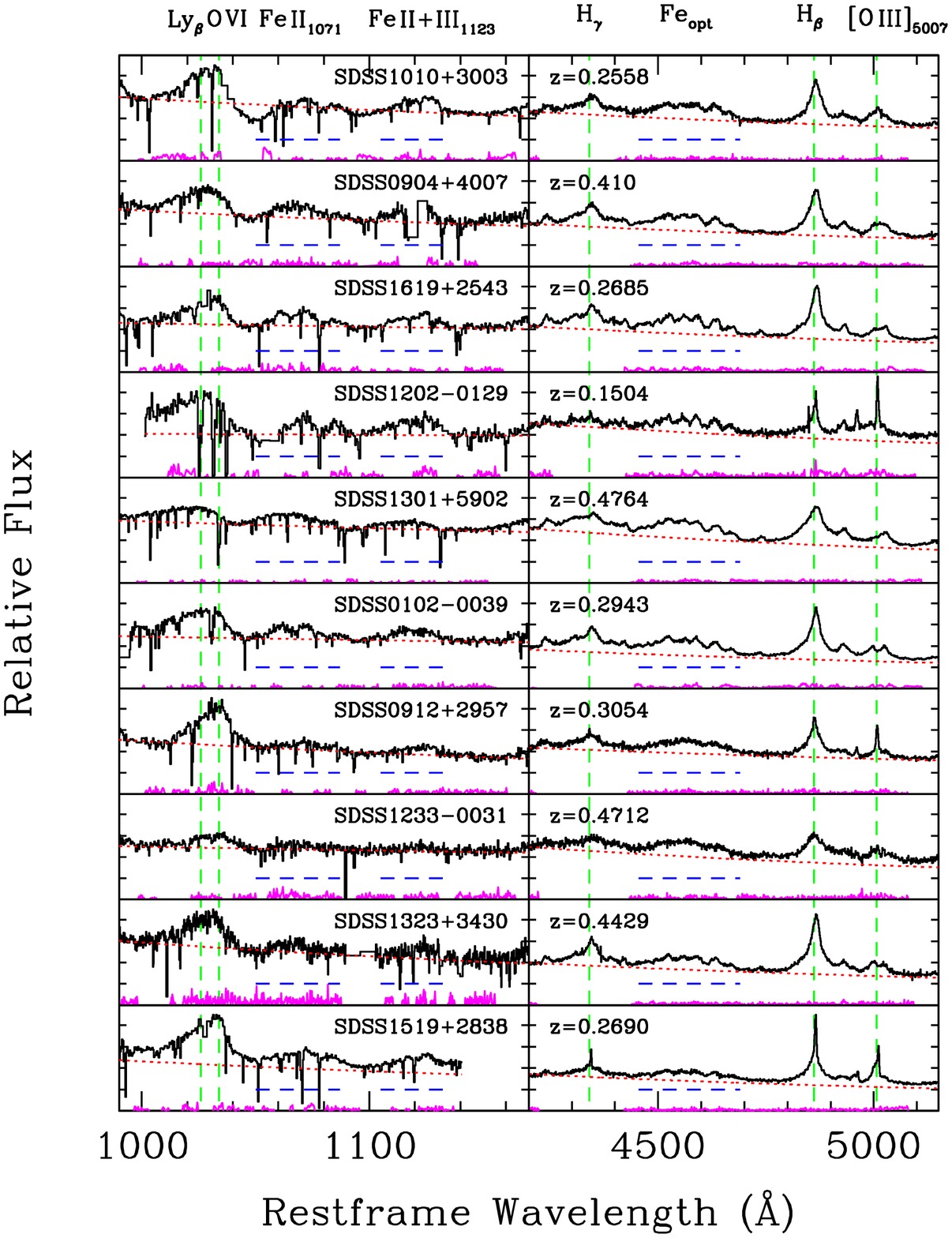}
        \caption{Spectra of ten SDSS quasars with the highest \ratio\ ratios 
          ($> 1.1$) or \feuv/(\lyb+\ion{O}{6}) ratios ($> 0.4$;
          see Table \ref{tbl-strong}). 
          Fluxes are scaled for proper display. Disconnected data points are instrumental
          gaps or the results of removing geocoronal contaminations.
          Ly$\beta$, \ovi, \hg, \hb, and \oiii\ are labeled at the top, 
          and their central wavelengths are marked with vertical green lines. 
          The broad \feuv\ and \feop\ features are labeled, and their 
          wavelength ranges are marked with horizontal blue lines. 
          The fitted power-law continua are plotted in red dotted lines, 
          and the absolute values of {\tt Specfit} residues in  the fitting windows 
          are plotted in magenta.
        \label{strong}}
\end{figure}

\begin{figure}
        \includegraphics[width=\columnwidth]{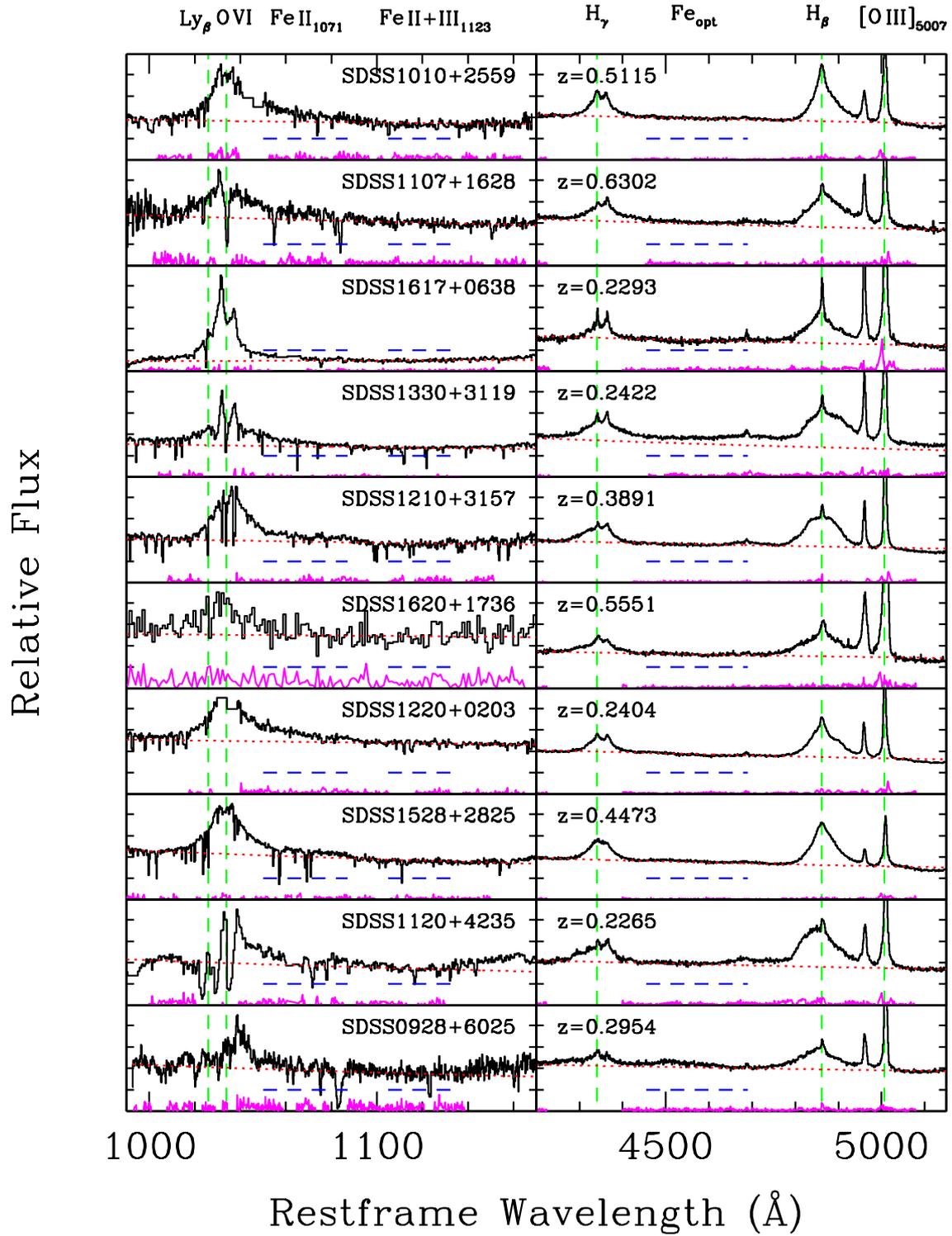}
        \caption{Spectra of ten SDSS quasars with the lowest \ratio\ ratios 
          $\le 0.11$. see Table \ref{tbl-weak}). 
          The same caption as Figure \ref{strong}.
          \label{weak}}
\end{figure}

\fe\ emission is weak and broad, and its measurements depend heavily on the 
accuracy of an underlying continuum. 
Unfortunately, there is no emission-free region in the vicinity of \hb. 
Two narrow windows at $4180-4220$ and $5080-5120$~\AA\ were chosen to provide a baseline 
of the underlying continuum. Overall, the optical fitting windows are $4180-4220$
and $4415-5120$~\AA, excluding \hg.
In the FUV band, the nominal fitting range was between 995 and 1170 \AA\, if data were available. 
The continuum windows free of emission lines were $995 - 1005$, $1090 - 1100$, and $1140 - 1150$ 
\AA, and at least two of them should be available for fitting. 
If the data extend to 1280 \AA, an additional window of $1270 - 1280$ \AA\ was 
added to improve the accuracy. 
\lya\ emission was fitted with dual Gaussian components along with dual \nv\
components. The centroid wavelengths of \nv\ components were tied to their 
respective \lya\ counterparts by the intrinsic atomic values.  
Additional complications arise from very broad red wings of \ovi\
that may extend beyond \uva. The \lyb\ component was a Gaussian with a 
centroid at $\sim 1026$ \AA, and \ovi\ with dual broad components: one
at FWHM $ 1500 - 3000$ \kms, and another at $> 3000$~\kms. 
The wavelength ratio of \lyb\ and the broader \ovi\ component was fixed based on 
their intrinsic atomic data, and the \lyb\ width was tied to that of the broader \lya\ component. 
The two FUV \fe\ features were each fit with a Gaussian of FWHM $\la 7000$ \kms. 
Figure \ref{exam} illustrates an example of  the optical and FUV spectra
of a \fe-strong quasar with the fitting components, residues and fitting windows.
Figure \ref{strong} displays the spectra of ten quasars with the highest \ratio\ or 
\feuv/(\lyb+\ion{O}{6}) ratios,
and Figure \ref{weak} for ten other quasars with the lowest \ratio\ ratios.

To calculate the FWHM and its error from a summed profile of multiple Gaussian components 
(\hb, \lya\ and \lyb+\ion{O}{6})
the widths of every component were generated randomly 200 times from its FWHM and error, and then
200 summed profiles were made. For each summed profile, its peak position was determined, 
and then the FWHM was measured at the half-peak positions. This procedure was repeated 200 times 
for each line, and the average and standard deviation of these measurements were tabulated as 
the FWHM and its error. 

The emission-line measurements in these selected quasars are listed in Tables \ref{tbl-strong} and 
\ref{tbl-weak}, respectively. The last column in these tables shows whether a quasar has 
been in the S06 sample. The errors in these tables were calculated from
the {\tt Specfit} results, and those of line ratios propagated through 
mathematical formulas for the merging components. 
To calculate the propagated errors for ratios between two variables, 
their relative errors were in an RMS relation. Namely, if A, $\Delta A$ and 
B, $\Delta B$ are the two variables with their respective errors, the errors of 
ratio C=(A/B) is interpreted as $\Delta C/C = \sqrt{(\Delta A/A)^2 + (\Delta B/B)^2}$. 
For \lyb+\ion{O}{6}, one \lyb\ and two \ion{O}{6} components were included, and, for \hb, 
\ion{O}{6} and \lya, two components were included.
Note that the actual errors are likely larger than the tabulated values, as
{\tt Specfit} values may not include all possible uncertainties, especially when
strong absorption is present near a line center.
In Tables \ref{tbl-strong2} and \ref{tbl-weak2}, the line widths of \ion{O}{6}+\lyb,  \lya, and \hb\ in the 
two groups are tabulated, respectively. 

\section{Discussion}\label{3}

\subsection{Relation between UV and optical \fe\ ratios}\label{sec_disc}

\begin{figure}
        \includegraphics[width=\columnwidth]{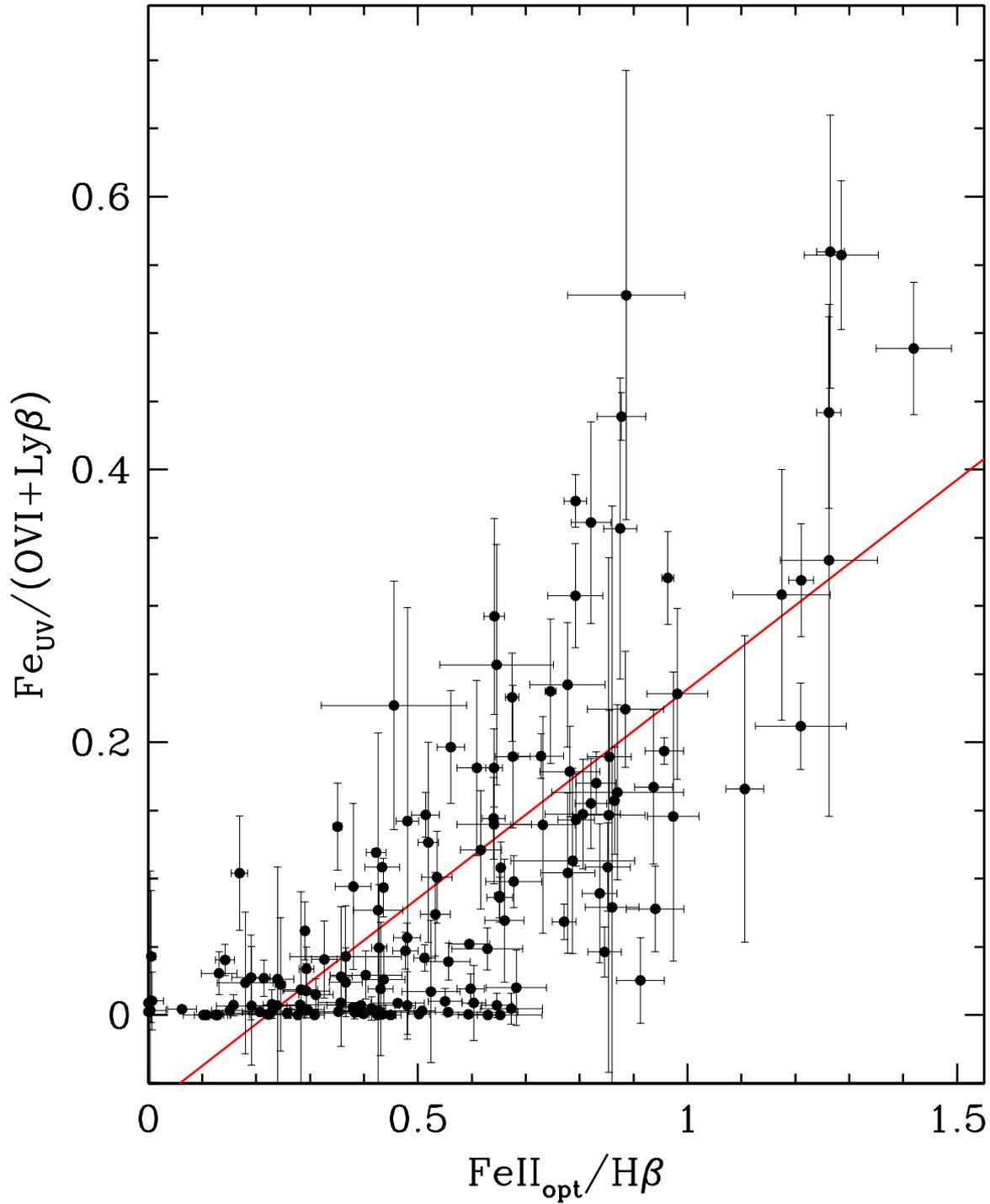}
        \caption{EW ratios of \feuv/(\lyb+\ion{O}{6}) vs. \ratio. 
          The Pearson coefficient of 0.75 suggests a significant
          correlation between them. 
The line of best linear regression 
          of slope 0.31 is plotted in red.
          The errors are propagated from the {\tt Specfit} fitting 
            results of individual Gaussian components.
        \label{uvb}}
\end{figure}

The \feuv\ emission is marked as \ion{Fe}{3} (UV 1) by \citet{laor,vb,harris}, 
and as \feuv\ by S06. Figure \ref{uvb} displays the \feuv/(\lyb+\ion{O}{6}) ratios 
vs. the \feop/\hb\ ratios for the sample.  
The Pearson coefficient of $R = 0.748$ suggests a significant correlation, 
with a probability (P-value) of $< 10^{-10}$ for an uncorrelated dataset 
producing such a distribution.
The linear regression slope is 0.307 and an intercept is $-0.068$.
To understand the effect of data dispersion to the correlation, the Pearson test was 
repeated 100 times with simulated data errors. In each run, the data points were modified
with errors, which were generated randomly under a normal distribution for the 
given standard deviations. The Pearson coefficients were between 0.58 and 0.75, with an 
average value of $\bar{R} = 0.68 \pm 0.04$. Even at the lowest value of $R =0.58$, the P-value 
is still smaller than $10^{-10}$. 

The trend suggests that quasars with higher \feuv\ emission likely show
high \feop, thus providing an epoxy of the Eddington ratios in high-redshift 
quasars where optical emission is not readily observable.
Note that the dispersion shown in Figure \ref{uvb} is significant; therefore,
a prediction of the Eddington ratio for a given quasar would still be uncertain.
At \ratio\ ratios $\la 0.7$, the \feuv\ features in many quasars are not detected. Some of these quasars with very low \ratio\ ratios may be referred as ``Pop B quasars'' \citep{marziani18}.
If the data points at  
\feuv/(\lyb+\ion{O}{6}) $ < 0.025 $ are excluded as their values are below the average errors in this section of data,
the remaining data of 90 quasars yield the Pearson coefficient of 0.706 
and the P-value at $< 10^{-10}$.
For a group of quasars with low \feuv/(\lyb+\ion{O}{6}) ratios ($<0.1$), their average \ratio\ 
ratio is $0.39 \pm 0.21$. Comparing with Figure 2 of \citet{marziani18}, these 
quasars are of type B1 and A1. For the group with high \feuv/(\lyb+\ion{O}{6}) 
ratios ($ > 0.3$), their \ratio\ ratio is $1.06 \pm 0.21$. These quasars are 
mostly of type A3. The Eddington ratios of these two groups, as 
judged from the optical \fe\ strengths, are noticeably different.

\begin{figure}
        \includegraphics[width=\columnwidth]{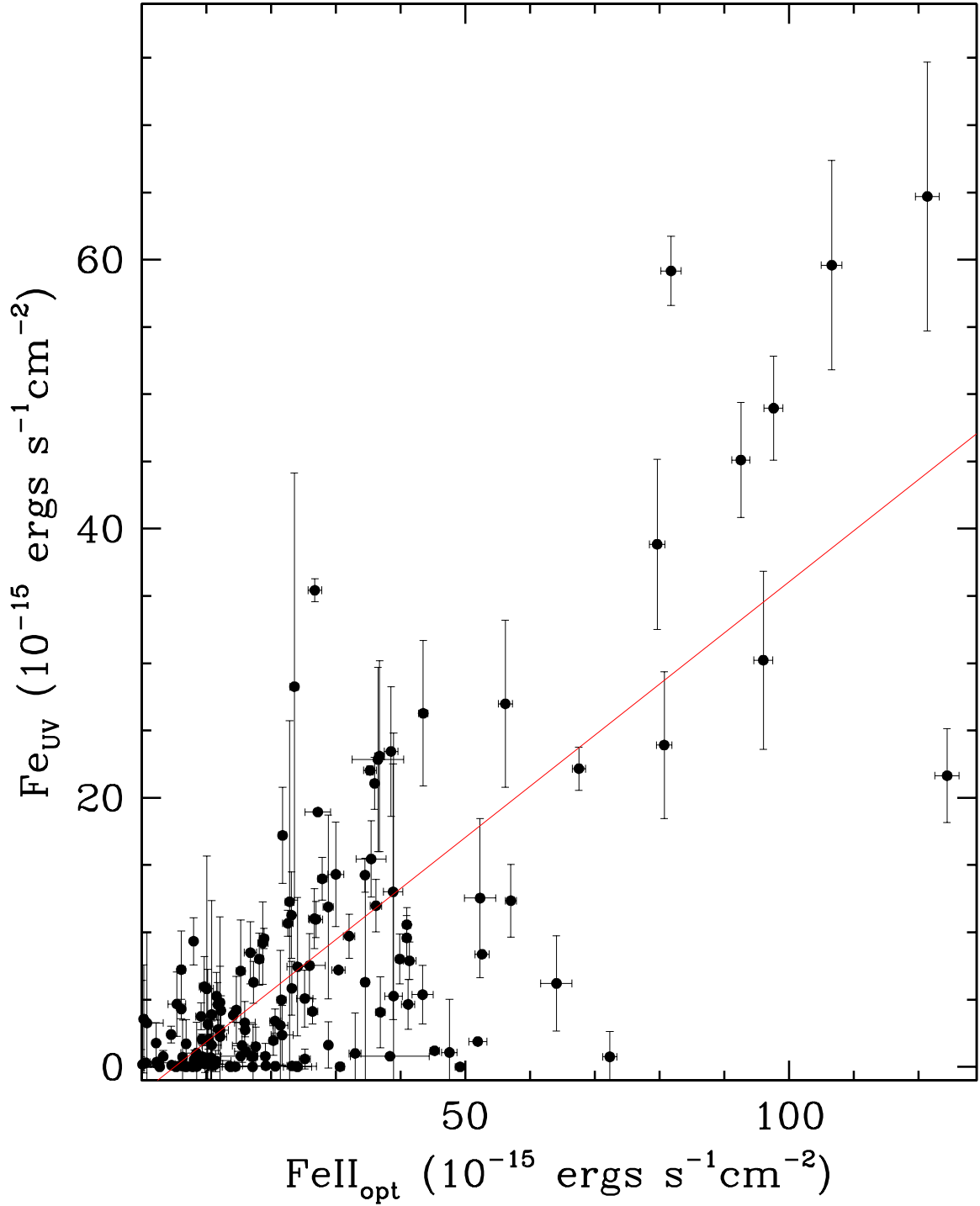}
        \caption{Restframe fluxes  of \feuv\ vs. \ion{Fe}{2}$_{\rm opt}$ emission. 
                    The Pearson coefficient of 0.76 suggests a significant correlation. 
          The line of best linear regression of slope 0.38 is plotted in red.
        \label{optuv}}
\end{figure}

Figure \ref{uvb} is plotted in line ratios, following the convention
in most publications on \fe\ properties.  Since both \HST\ and SDSS spectra are 
photometric \citep{hirsch,ivezic}, it would be interesting to directly compare the
FUV and optical \fe\ fluxes, as shown in Figure \ref{optuv}.
While the data display significant dispersion, the Pearson coefficient between
the \feuv\ and \feop\ fluxes is 0.763,
and their linear relationcan be described as \ \ \ \feuv\ = 0.381 \feop\ $-\ 1.9$ \ \ \ 
where fluxes are in units of $10^{-15}$ \flux. The corresponding P-value is 
smaller than $ 10^{-10}$.
The tests with simulated data errors yielded an average value of $\bar{R} = 0.73 \pm 0.02$, 
corresponding to P-values smaller than $10^{-10}$. 
The trend in this figure supports that the strengths of \feuv\ and \feop\ are indeed correlated.

\begin{figure}
        \includegraphics[width=\columnwidth]{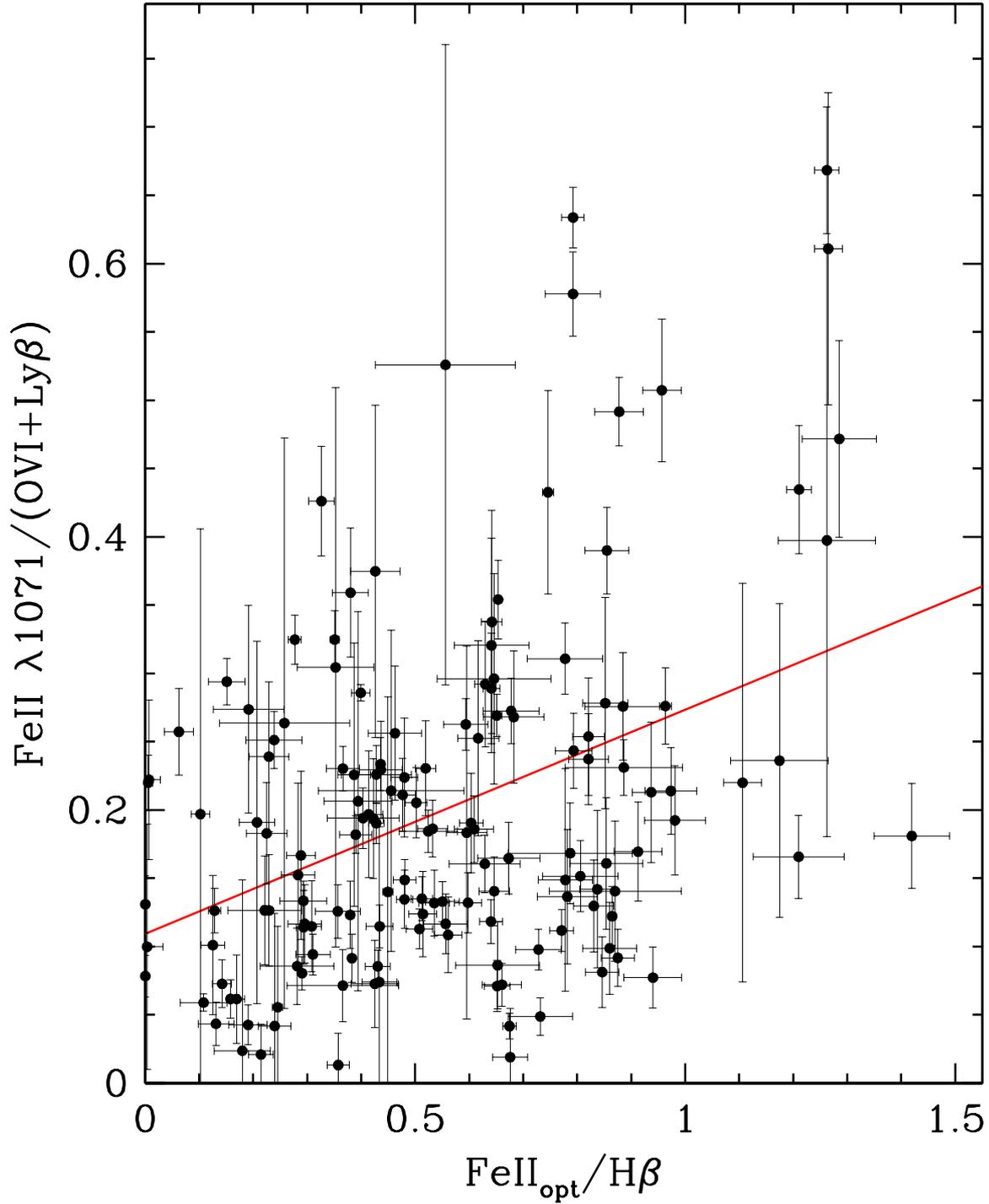}
        \caption{EW ratios of \uva/(\lyb+\ion{O}{6}) vs. \ratio. 
           The Pearson coefficient of 0.40 suggests a moderate 
           correlation between them. The best linear regression 
          of slope of 0.17 is plotted in red.  See caption of Figure \ref{uvb}.
        \label{uva}}
\end{figure}

The emission feature around 1074 \AA\ was identified as \ion{Ar}{4} \citep{vb},
as "unknown'' by \citet{harris}, and as \uva\ by S06. 
Between \uva/(\lyb+\ion{O}{6}) and \feop/\hb, 
the Pearson coefficient is 0.395, and the P-value is $< 6 \times 10^{-7}$.
The regression line slope is 0.164 and the intercept 0.11.
Since there are no other Ar lines and since the strength of this feature displays a 
moderate correlation with \feop, it is likely a \fe\ feature. 
The lower significance for \uva\ is probably attributed to contaminations 
from the red broad \ovi\ wing (Figure \ref{uva}). 

\subsection{Properties of other FUV lines}

\begin{figure}
        \includegraphics[width=\columnwidth]{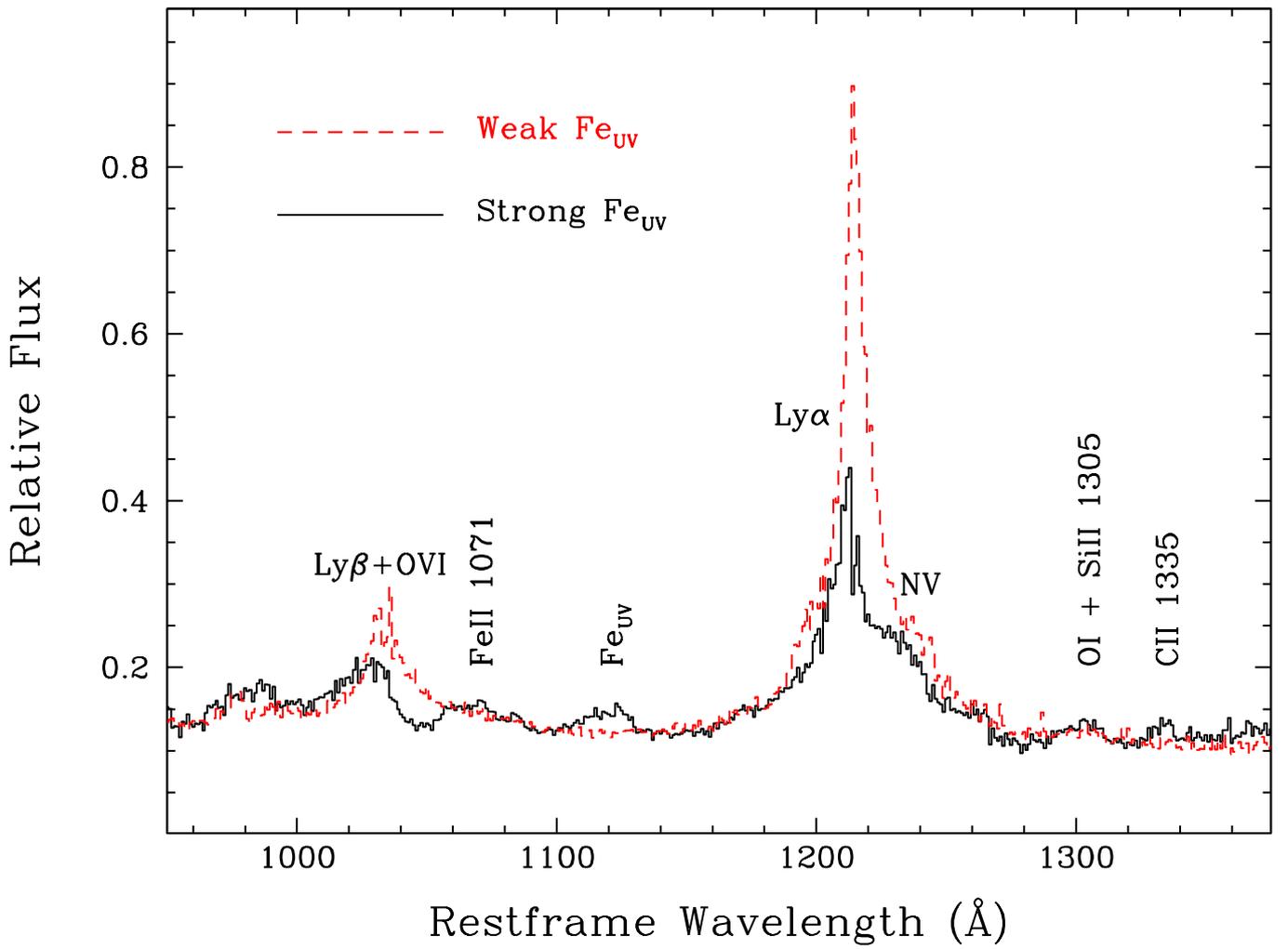}
        \caption{Summed spectra of quasars with different \feuv\ strengths. The spectrum in black 
          is that of the ten strongest \fe\ (Figure \ref{strong} and Table \ref{tbl-strong}), and that 
          in red dashes of the ten weakest \fe\ (Figure \ref{weak} and Table \ref{tbl-weak}).  
          Note the difference in \cii.         \label{comp}}
\end{figure}

The spectra of \fe-strong quasars in Table \ref{tbl-strong} and Figure \ref{strong}
were combined after normalizing with the fluxes around 1100~\AA.
Similarly, a combined spectrum was made for ten \fe-weak quasars in Table \ref{tbl-weak}
and Figure \ref{weak}. As shown in Figure \ref{comp}, the \lya\ emission in \fe-weak quasars displays 
a stronger narrow component.
A {\tt Specfit} fit with dual components for \lya\ and \ion{O}{6}+\lyb\ and single
components for other emission lines was carried out.
The EWs of narrow \lya\ components are $ 11.6 \pm 1.3$ 
\AA\ for \fe-strong quasars and 
$ 34.0 \pm 2.5 $ \AA\ for \fe-weak quasars. The respective \feuv\ EWs are 
$ 4.5 \pm 0.7 $ and  $0.3 \pm 1.0 $ \AA.
To check if the trend is a small-sample fluctuation, the same test was carried 
out for 20 \fe-strong quasars and other 20 \fe-weak ones. The EW of narrow \lya\ components 
are $18.1 \pm 1.3 $ and $ 33.0 \pm 1.8$ \AA, respectively.  
Since there is a minimal overlap between these quasars and those in S06, this trend provides
additional support to the results in S06. It suggests that Class-I
quasars are \fe\ weak, \lya\ strong and have low Eddington ratios, 
in contrast to their Class-III counterparts.

\citet{sulentic02} studied the spectral features for different groups in the diagram of \ratio\ ratios 
vs. \hb\ FWHM. These groups display distinct \feop\ strengths as well as \hb\ widths. 
Based on the correlation between \feuv\ and \feop\ (Figures \ref{uvb} and \ref{optuv}), it may be possible to link 
the UV classes of S06 with the optical groups of \citet{sulentic02}: 
Class-I quasars show absent \feuv\ and are hence related to the B-type in the optical. Class-III quasars show 
significant \feuv\ and are related to the A-3 type.

It is also noted in Figure \ref{comp} that the \cii\ emission is correlated with \fe\ strengths. 
Between the ratios of \cii/\lya\ and \ratio\ ratios of 51 objects, the Pearson coefficient 
of  0.475 and the P-value of 0.04\% suggest a moderate correlation. 

Note that both the \feuv\ and \cii\ emission is of low ionization, and their strengths are 
different from major FUV emission lines, which are of high ionization. Emission lines of different 
ionization levels are believed to originate in separated regions 
\citep{shang07,richards11,bddt}. The results in this study suggest that some 
low-ionization lines may scale Eddington ratios.

\begin{figure}
        \includegraphics[width=\columnwidth]{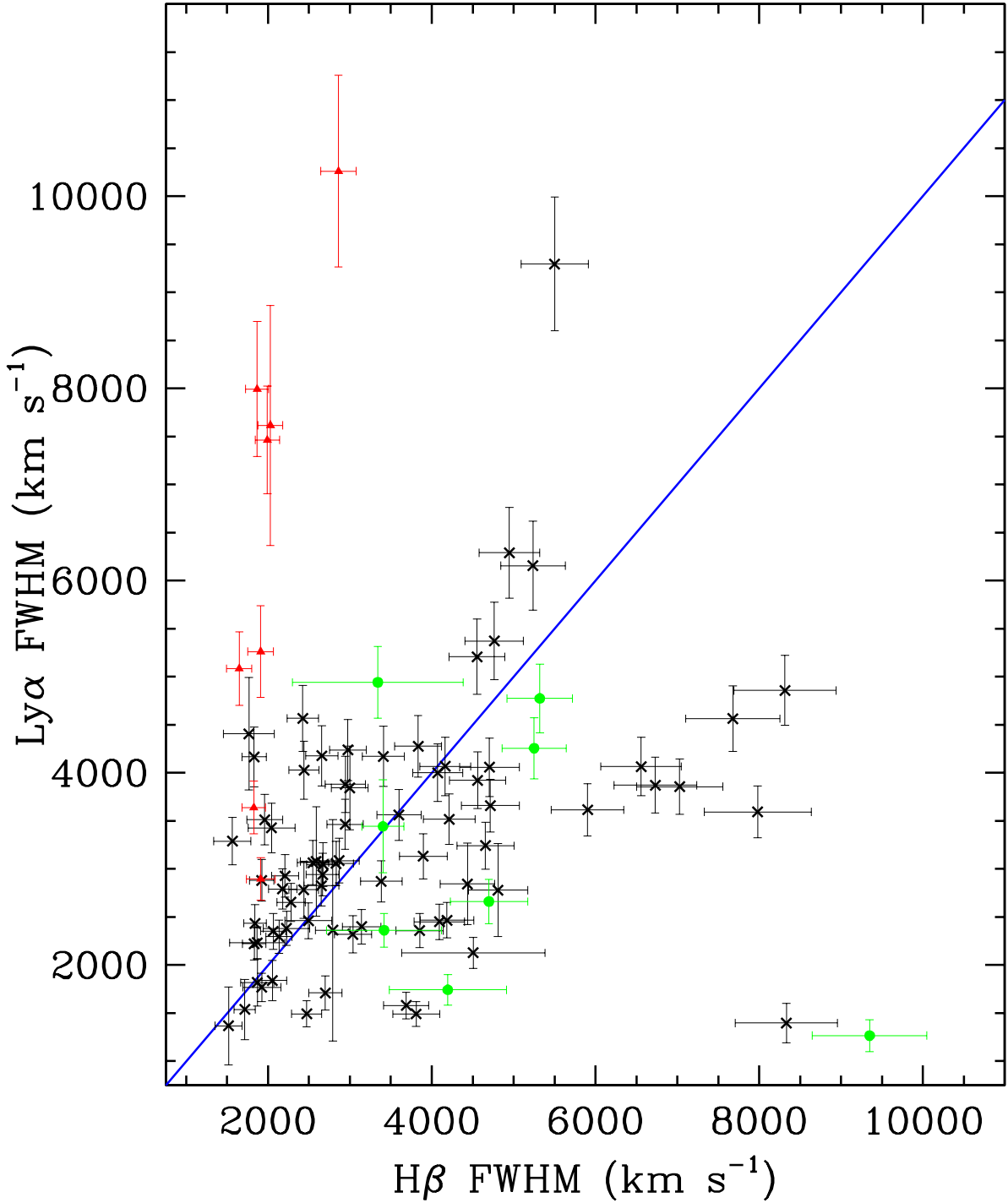}
        \caption{FWHMs of \lya\ vs. \hb. The blue line marks an ideal situation of equal FWHMs. 
          The Pearson coefficient of 0.08 suggests a poor correlation. 
          The distributions of \fe-strong (red) and \fe-weak (green) quasars 
          are noticeably different.
        \label{width}}
\end{figure}

\mgii\ emission is of low ionization; therefore, it is conceivable that the FWHMs of \mgii\ are 
similar to those of \hb\ \citep{mclure,wang}, and they were used to estimate the central 
black-hole masses. \citet{skc} studied near-UV (NUV) lines and compared them with \ratio. They 
found that the \mgii\ properties bear resemblance with \hb, and the NUV \fe\ strengths are correlated to \feop.
However, major FUV lines are of high ionization. Attempts have been made to explore the 
possibility of using FUV emission-line widths to scale the accretion rate, and the results remain
unconvincing \citep{shemmer,martinez, sulentic17}, as the \civ\ widths are significantly different 
from those of \hb. As shown in Figure \ref{width}, the FWHM of \lya\ and \hb\ are uncorrelated, with 
the Pearson coefficient of 0.079 and the  P-value of 46\%  (90 data points). 

The \fe-strong quasars, as shown by red symbols, have narrow \hb\ widths and randomly 
distributed \lya\ widths. The \fe-weak quasars, as shown in green symbols, have narrow \lya\ widths and moderate-to-broad \hb\ widths.
Given the significant dispersion, the central masses and accretion rates derived from \lya\ 
would be inconsistent with those from \hb. 

\begin{figure}
        \includegraphics[width=\columnwidth]{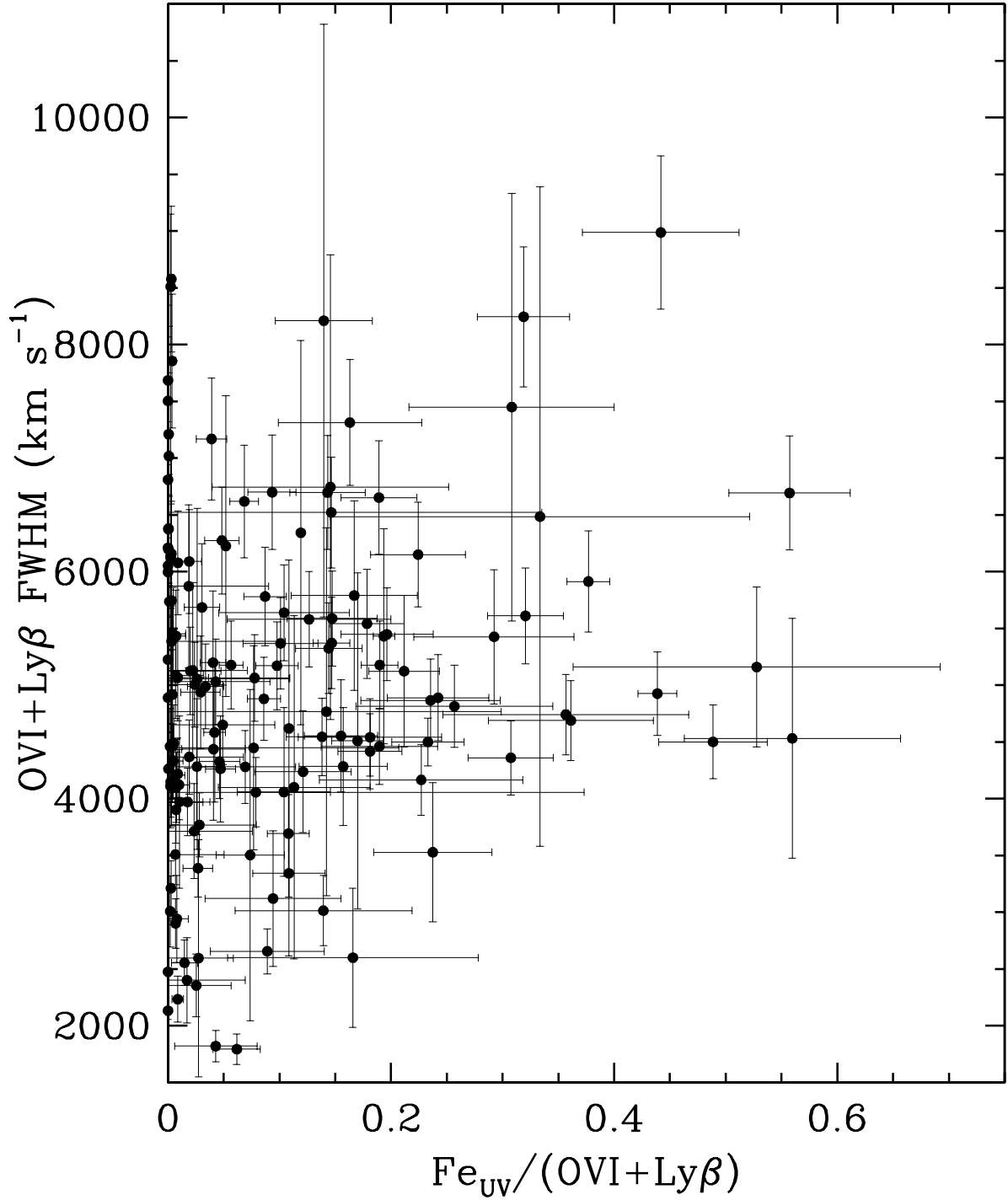}
        \caption{
          Ratios of \feuv/(\lyb+\ion{O}{6}) vs. line widths of \lyb+\ion{O}{6}. 
          Quasars with strong \fe\ features tend to display broader emission, contrary to 
          a trend in the optical band. 
        \label{o6}}
\end{figure}

It has been suggested \citep{shen14,marziani18} that strong \ratio\ ratios 
are related to high accretion rates. 
It is also known that strong \feop\ is associated with narrow \hb\ \citep{sulentic02}.
The nature of a broad span of \hb\ line widths is complex: while they are 
attributed to an orientation effect, other effects may also be at work.
In virialized cases, the \hb\ line widths are also related to the black-hole masses
and accretion rates \citep{sulentic17}.
In other words, there appear to be intrinsic difference 
between Type-A and B quasars in Figure 2 of \citet{marziani18}.
In Figure \ref{o6}, the widths of \ion{O}{6}+\lyb\ are plotted vs. the ratios of \ratio.  
Strong \feuv\ seems to be associated with broader \ion{O}{6}+\lyb.
This trend illustrates the difficulty of using the widths of \lya\ and other FUV lines \citep{richards11}
to estimate the accretion rates. For this sample, the Pearson efficient between the widths of 
\ion{O}{6}+\lyb\ and \hb\ is merely 0.065 (P-value 43\%).

\subsection{Redshift effect}
Between redshifts $ 0.1 - 0.7$, the cosmic distances span a large range by a factor of $\sim 5$, 
and this quasar sample covers the absolute luminosities between $-21.3$ and $-26.4$. 
To check if the observed trend is redshift dependent, a statistical test
was carried out between the \ratio\ ratios and redshifts. 
It yielded the Pearson coefficient of 0.064 and the P-value of 43\%, suggesting a null correlation. 

The correlation between \ratio\ and the absolute magnitude is also null, with a 
Pearson coefficient of $-0.023$ and the P-value of 77\%. 
Note that the average redshift in the \fe-weak group (Table \ref{tbl-weak}) is slightly higher than 
that in the \fe-strong group (Table \ref{tbl-strong}).
Overall, the correlation shown in Figures \ref{uvb} and \ref{optuv}
is not attributed to redshifts or luminosities.

\section{Summary}\label{sum}
Based on a large database of FUV and optical spectra of low-redshift quasars, we find that
(1) the relation between FUV and optical emission lines is complex: 
the line widths of \lya\ are uncorrelated with those of \hb, making it difficult to use 
\lya\ to estimate the masses of central black holes;
(2) the equivalent widths of high- and low-ionization lines exhibit opposite correlations with \feop;
(3) the significant correlation between \feuv\ and \feop\ strengths makes it possible 
to use the FUV \fe\ emission as a proxy of the Eddington ratios; and 
(4) the quasars with high Eddington ratios tend to display broader \lya\ and \ion{O}{6} emission, 
contrary to a trend in  the optical band.

A proxy of the Eddington ratios would be valuable at higher redshifts, when 
\feop\ and \hb\ emission features are redshifted out of the optical band.
The results suggest that the quasar classification of S06 in the FUV band
is driven by the Eddington ratios, consistent with
previous results of the optical and X-ray bands.

\section*{acknowledgments}

The author thanks the anonymous referee for many thoughtful comments.

Funding for the SDSS and SDSS-II has been provided by the Alfred P. Sloan Foundation, 
the Participating Institutions, the National Science Foundation, the U.S. Department 
of Energy, the National Aeronautics and Space Administration, the Japanese 
Monbukagakusho, the Max Planck Society, and the Higher Education Funding Council for 
England. The SDSS Web Site is http://www.sdss.org/.

This research has made use of the NASA/IPAC Extragalactic Database (NED),
which is operated by the Jet Propulsion Laboratory, California Institute of Technology,
under contract with the National Aeronautics and Space Administration (NASA). 
It is based on observations made with the NASA/ESA Hubble Space Telescope, obtained from the data archive at 
the Space Telescope Science Institute. STScI is operated by the Association of Universities for Research 
in Astronomy, Inc. under NASA contract NAS 5-26555.
Some of the data presented in this paper were obtained from the Mikulski Archive for Space
Telescope (MAST). STScI is operated by the Association of Universities for Research in 
Astronomy, Inc., under NASA contract NAS5-26555. 

\section*{Data Availability Statement}
All the data and software in this report are public.

\begin{landscape}
\begin{table*}  
	\centering
        \caption{Emission Line Strengths in \fe-Strong Quasars} 
        \label{tbl-strong}
        \begin{tabular}{cc|cccccc|cc|c} 
        \hline
\multicolumn{2}{c|}{Quasar} &
\multicolumn{6}{c|}{Equivalent Width (\AA)} & \multicolumn{2}{c|}{Flux Ratio} & S06
\\  
\cmidrule(lr){1-2}\cmidrule(lr){3-8}\cmidrule(lr){9-10}
Name & Redshift & \fe & \hb & \feuv & \lyb+\ion{O}{6} & \lya & \ion{C}{2} & \feop/\hb & \feuv/(\lyb+\ion{O}{6})  & 
\\ 
\multicolumn{2}{c}{} &
\lm 4565 &
\lm 4861 & 
\lm 1123 &
\lm\lm 1026,1034 & \lm 1216 & \lm 1335 & & & \\
\hline
SDSS1010+3003 & 0.2558 & $68.8 \pm 0.7 $& $48.5 \pm 0.1 $& $5.0 \pm 0.1 $& $10.2 \pm 0.1 $ & $53 \pm 3 $ & $2.1 \pm 0.5 $ & $1.42 \pm 0.06 $& $0.49 \pm 0.05 $ &  NO \\ 
SDSS0904+4007 & 0.4100 & $78.9 \pm 2.1 $& $61.4 \pm 0.1 $& $6.0 \pm 0.1 $& $10.8 \pm 0.1 $ & $95 \pm 4 $ &  --- & $1.28 \pm 0.07 $& $0.56 \pm 0.05 $ &  NO \\ 
SDSS1619+2543 & 0.2685 & $81.2 \pm 0.5 $& $64.2 \pm 0.1 $& $5.7 \pm 0.1 $& $10.2 \pm 0.2 $ & $87 \pm 2 $ & $3.2 \pm 0.8 $ & $1.26 \pm 0.02 $& $0.56 \pm 0.10 $ &  NO \\ 
SDSS1202$-$0129 & 0.1504 & $50.6 \pm 1.4 $& $40.1 \pm 0.1 $& $7.4 \pm 0.6 $& $22.1 \pm 0.1 $ & $42 \pm 2 $ &  --- & $1.26 \pm 0.09 $& $0.33 \pm 0.19 $ &  NO \\ 
SDSS1301+5902 & 0.4764 & $74.9 \pm 0.5 $& $59.4 \pm 0.1 $& $3.5 \pm 0.2 $& $7.9 \pm 0.1 $ & $56 \pm 5 $ &  --- & $1.26 \pm 0.02 $& $0.44 \pm 0.07 $ &  YES \\ 
SDSS0102$-$0039 & 0.2943 & $90.5 \pm 0.7 $& $74.8 \pm 0.1 $& $5.0 \pm 0.1 $& $15.8 \pm 0.1 $ & $61 \pm 5 $ & $2.6 \pm 3.0 $ & $1.21 \pm 0.02 $& $0.32 \pm 0.04 $ &  NO \\ 
SDSS0912+2957 & 0.3054 & $50.2 \pm 1.7 $& $41.5 \pm 0.1 $& $2.9 \pm 0.1 $& $13.9 \pm 0.1 $ & $87 \pm 2 $ & $13.9 \pm 1.2 $ & $1.21 \pm 0.08 $& $0.21 \pm 0.03 $ &  NO \\ 
SDSS1233$-$0031 & 0.4712 & $64.8 \pm 2.1 $& $55.2 \pm 0.1 $& $1.9 \pm 0.2 $& $6.1 \pm 0.2 $ &  --- &  --- & $1.17 \pm 0.09 $& $0.31 \pm 0.09 $ &  NO \\ 
SDSS1323+3430 & 0.4429 & $67.0 \pm 7.3 $& $75.6 \pm 0.1 $& $5.1 \pm 0.3 $& $9.7 \pm 0.1 $ & $86 \pm 5 $ &  --- & $0.89 \pm 0.11 $& $0.53 \pm 0.16 $ &  NO \\ 
SDSS1519+2838 & 0.2690 & $55.2 \pm 2.2 $& $62.9 \pm 0.1 $& $12.4 \pm 0.1 $& $28.3 \pm 0.1 $ &  --- &  --- & $0.88 \pm 0.04 $& $0.44 \pm 0.02 $ &  NO \\ 
\hline 
 Average /Deviation &  $0.33 \pm 0.10 $ &$68.2 \pm 12.8 $&$58 \pm 12 $&$5.5 \pm 2.8 $&$13 \pm 7 $&$71 \pm 19 $&$5.5 \pm 4.9 $&$1.19 \pm 0.16 $&$0.42 \pm 0.11 $& \\ \hline
\end{tabular}
\end{table*} 
\begin{table*}  
	\centering
        \caption{Emission Line Widths in \fe-Strong Quasars} 
        \label{tbl-strong2}
        \begin{tabular}{c|ccc} 
        \hline
Quasar & \multicolumn{3}{c|}{FWHM (\kms)} 
\\  
\cmidrule(lr){2-4}
 & \lyb+\ion{O}{6}&\lya & \hb
\\ 
 & \lm\lm 1026,1034 & \lm 1216 & \lm 4861\\
\hline
SDSS1010+3003 & $4500 \pm 325 $ & $5262 \pm 476 $ & $1909 \pm  157 $    \\ 
SDSS0904+4007 & $6693 \pm 502 $ & $7616 \pm 1249 $ & $2028 \pm  152 $    \\ 
SDSS1619+2543 & $4531 \pm 1056 $ & $3639 \pm 273 $ & $1826 \pm  143 $    \\ 
SDSS1202$-$0129 & $6485 \pm 2906 $ & $7463 \pm 560 $ & $1992 \pm  149 $    \\ 
SDSS1301+5902 & $8987 \pm 674 $ & $10260 \pm 997 $ & $2860 \pm  215 $    \\ 
SDSS0102$-$0039 & $8244 \pm 618 $ & $7993 \pm 703 $ & $1866 \pm  140 $    \\ 
SDSS0912+2957 & $5123 \pm 668 $ & $2895 \pm 220 $ & $1911 \pm  176 $    \\ 
SDSS1233$-$0031 & $7449 \pm 1882 $ &  --- & $3678 \pm  277 $    \\ 
SDSS1323+3430 & $5160 \pm 704 $ & $5085 \pm 381 $ & $1649 \pm  154 $    \\ 
SDSS1519+2838 & $4925 \pm 369 $ &  --- & $2074 \pm  156 $    \\ 
\hline 
 Average /Deviation & $6210 \pm 1529 $&$6277 \pm 2315 $ &$2179 \pm 585 $  \\ \hline
\end{tabular}
\end{table*} 
\end{landscape}                 

\begin{landscape}
\begin{table*}
        \centering
        \caption{Emission Line Strengths in \fe-Weak Quasars} 
        \label{tbl-weak}
        \begin{tabular}{cc|cccccc|cc|c} 
        \hline
\multicolumn{2}{c}{Quasar} &
\multicolumn{6}{c}{Equivalent Width (\AA)} & \multicolumn{2}{c}{Flux Ratio} & S06
\\ 
\cmidrule(lr){1-2}\cmidrule(lr){3-8}\cmidrule(lr){9-10}
Name & Redshift & \fe & \hb & \feuv & \lyb+\ion{O}{6} & \lya & \ion{C}{2} & \feop/\hb & \feuv/(\lyb+\ion{O}{6}) & 
\\ 
& &\lm 4565 &
\lm 4861 & 
\lm 1123 &
\lm\lm 1026,1034 & \lm 1216 & \lm 1335 &
\multicolumn{2}{c}{} & \\
\hline
SDSS1010+2559 & 0.5115 & $4.8 \pm 2.1 $& $76.8 \pm 0.1 $& $0.1 \pm 0.1 $& $24.0 \pm 0.1 $ &  --- &  --- & $0.06 \pm 0.03 $& $0.00 \pm 0.01 $ &  NO \\ 
SDSS1107+1628 & 0.6302 & $0.0 \pm 0.1 $& $98.2 \pm 0.1 $& $0.0 \pm 1.2 $& $16.7 \pm 0.1 $ & $93 \pm 8 $ & $0.0 \pm 0.1 $ & $0.00 \pm 0.01 $& $0.00 \pm 0.01 $ &  YES \\ 
SDSS1617+0638 & 0.2293 & $0.0 \pm 0.1 $& $102.5 \pm 0.1 $& $0.9 \pm 0.6 $& $98.7 \pm 0.1 $ & $512 \pm 6 $ & $0.1 \pm 1.1 $ & $0.00 \pm 0.01 $& $0.01 \pm 0.01 $ &  NO \\ 
SDSS1330+3119 & 0.2422 & $37.1 \pm 3.7 $& $173.3 \pm 0.1 $& $0.8 \pm 0.5 $& $31.1 \pm 0.1 $ & $106 \pm 2 $ & $0.4 \pm 0.5 $ & $0.21 \pm 0.02 $& $0.03 \pm 0.01 $ &  NO \\ 
SDSS1210+3157 & 0.3891 & $0.5 \pm 0.7 $& $81.7 \pm 0.1 $& $0.9 \pm 1.1 $& $20.2 \pm 0.1 $ & $173 \pm 4 $ &  --- & $0.01 \pm 0.01 $& $0.04 \pm 0.05 $ &  NO \\ 
SDSS1620+1736 & 0.5551 & $0.4 \pm 3.0 $& $99.8 \pm 0.1 $& $0.0 \pm 29.7 $& $8.7 \pm 0.2 $ & $96 \pm 6 $ & $0.1 \pm 0.1 $ & $0.00 \pm 0.03 $& $0.00 \pm 0.10 $ &  YES \\ 
SDSS1220+0203 & 0.2404 & $0.5 \pm 1.5 $& $72.7 \pm 0.1 $& $0.2 \pm 2.0 $& $20.4 \pm 0.1 $ & $99 \pm 5 $ &  --- & $0.01 \pm 0.02 $& $0.01 \pm 0.02 $ &  NO \\ 
SDSS1528+2825 & 0.4473 & $8.3 \pm 3.3 $& $77.1 \pm 0.1 $& $0.0 \pm 0.1 $& $21.8 \pm 0.1 $ & $133 \pm 19 $ &  --- & $0.11 \pm 0.04 $& $0.00 \pm 0.01 $ &  NO \\ 
SDSS1120+4235 & 0.2265 & $11.3 \pm 1.9 $& $110.7 \pm 0.1 $& $0.0 \pm 0.1 $& $20.0 \pm 0.1 $ &  --- &  --- & $0.10 \pm 0.02 $& $0.00 \pm 0.01 $ &  NO \\ 
SDSS0928+6025 & 0.2954 & $19.2 \pm 1.3 $& $50.6 \pm 0.1 $& $1.2 \pm 0.6 $& $12.8 \pm 0.1 $ & $90 \pm 29 $ & $2.1 \pm 0.8 $ & $0.38 \pm 0.03 $& $0.09 \pm 0.06 $ &  NO \\ 
\hline 
 Average /Deviation &  $0.38 \pm 0.14 $ &$8.2 \pm 11.4 $&$94 \pm 31 $&$0.4 \pm 0.4 $&$27 \pm 24 $&$163 \pm 135 $&$0.6 \pm 0.8 $&$0.09 \pm 0.12 $&$0.02 \pm 0.03 $& \\ \hline
\end{tabular}
\end{table*}

\begin{table*}
        \centering
        \caption{Emission Line Widths in \fe-Weak Quasars} 
        \label{tbl-weak2}
        \begin{tabular}{c|ccc} 
        \hline
Quasar & \multicolumn{3}{c}{FWHM (\kms)}
\\ 
\cmidrule(lr){2-4}
  & \lyb+\ion{O}{6} & \lya & \hb
\\ 
 & \lm\lm 1026,1034 &\lm 1216 & \lm 4681\\
\hline
SDSS1010+2559 & $4916 \pm 859 $ &  --- & $2411 \pm  181 $    \\ 
SDSS1107+1628 & $8510 \pm 638 $ & $4774 \pm 358 $ & $5320 \pm  399 $    \\ 
SDSS1617+0638 & $2234 \pm 202 $ & $1742 \pm 157 $ & $4196 \pm  717 $    \\ 
SDSS1330+3119 & $3387 \pm 254 $ & $1266 \pm 165 $ & $9349 \pm  701 $    \\ 
SDSS1210+3157 & $5030 \pm 377 $ & $4256 \pm 319 $ & $5251 \pm  394 $    \\ 
SDSS1620+1736 & $5385 \pm 1624 $ & $4942 \pm 371 $ & $3341 \pm  1045 $    \\ 
SDSS1220+0203 & $3970 \pm 760 $ & $2362 \pm 177 $ & $3417 \pm  701 $    \\ 
SDSS1528+2825 & $4890 \pm 367 $ & $3443 \pm 484 $ & $3403 \pm  255 $    \\ 
SDSS1120+4235 & $2133 \pm 796 $ &  --- & $4919 \pm  369 $    \\ 
SDSS0928+6025 & $3120 \pm 596 $ & $2661 \pm 232 $ & $4698 \pm  473 $    \\ 
\hline 
 Average /Deviation & $4358 \pm 1773 $&$3181 \pm 1301 $ &$4630 \pm 1816 $  \\ 
\hline
\hline
\end{tabular} 
\end{table*}
\bsp	\label{lastpage}
\end{landscape}

\end{document}